\documentclass[twocolumn,amssymb,amsmath,aps,prb,showpacs]{revtex4-2}
\usepackage{array}
\usepackage{graphicx}
\usepackage{color}
\usepackage{multirow}
\usepackage{float}
\usepackage{amssymb}
\usepackage{amsmath}
\usepackage{makecell}
\begin{document}
	\title{Two-band conduction as a pathway to non-linear Hall effect and unsaturated negative magnetoresistance  in the martensitic compound GdPd$_2$Bi}
	\author{Snehashish Chatterjee}
	\author{Saurav Giri} 
	\author{Subham Majumdar} 
	\email{sspsm2@iacs.res.in}
	\affiliation{School of Physical Sciences, Indian Association for the Cultivation of Science, 2A \& B Raja S. C. Mullick Road, Jadavpur, Kolkata 700032, India}
	\author{Prabir Dutta}
	\affiliation{Jawaharlal Nehru Centre for Advanced Scientific Research,Jakkur, Bangalore, 560064, India}
	\author {Surasree Sadhukhan}
	\author{Sudipta Kanungo}
	\email{sudipta@iitgoa.ac.in}
	\affiliation{School of Physical Sciences, Indian Institute of Technology Goa, Ponda 403401, Goa, India}
	\author{Souvik Chatterjee}
	\affiliation{UGC-DAE Consortium for Scientific Research, Kolkata Centre, Sector III, LB-8, Salt Lake, Kolkata 700106, India}
	\author {Manju Mishra Patidar}
	\altaffiliation{Presently at Emerald Heights International School, A.B. Road, Rau, Indore 453331, India}
	\author{Gunadhor Singh Okram}
	\author{V. Ganesan}
	\altaffiliation{Presently at Medi-Caps University, A. B. Road, Pigdamber, Rau, Indore 453331, India}
	\affiliation{UGC-DAE Consortium for Scientific Research, University Campus, Khandwa Road, Indore 452017, India}
	\author{G. Das}
	\altaffiliation{Presently at Elettra Sincrotrone Trieste, Strada Statale 14, km 163.5 in AREA Science Park, Basovizza, Trieste 34149, Italy}
	\author{V. Rajaji}
	\affiliation{Chemistry and Physics of Materials Unit, Jawaharlal Nehru Centre for Advanced Scientific Research, Jakkar, Bangalore, 560 064, India}

	\begin{abstract}
 The present work aims to address the electronic and magnetic properties of the intermetallic compound  GdPd$_2$Bi through a comprehensive study of the structural, magnetic, electrical and thermal transport on a polycrystalline sample, followed by theoretical calculations. Our findings indicate that the magnetic ground state is antiferromagnetic in nature. Magnetotransport data present prominent hysteresis loop hinting a structural transition with further support from specific heat and thermopower measurements, but no such signature is observed in the magnetization study. Temperature dependent powder x-ray diffraction measurements confirm martensitic transition from the high-temperature (HT) cubic Heusler $L2_1$ structure to the low-temperature (LT) orthorhombic $Pmma$ structure similar to many previously reported shape memory alloys. The HT to LT phase transition is characterized by a sharp increase in resistivity associated with prominent thermal hysteresis.  Further, we observe robust Bain distortion between cubic and orthorhombic lattice parameters related by $a_{orth} = \sqrt{2}a_{cub}$, $b_{orth} = a_{cub}$ and $c_{orth} = a_{cub}/\sqrt{2}$, that occurs by contraction along $c$-axis and elongation along $a$-axis respectively. The sample shows an  unusual  `non-saturating' $H^2$-dependent negative magnetoresistance for magnetic field as high as 150 kOe. In addition, non-linear field dependence of Hall resistivity is observed below about 30 K, which coincides with the sign change of the Seebeck coefficient.  The  electronic structure calculations confirm robust metallic states both in the LT and  HT phases. It indicates complex nature of the Fermi surface  along with the existence of both electron and hole charge carriers. The anomalous transport behaviors  can be related to the  presence of both electron and hole pockets.
	\end{abstract}
	\maketitle
	
\section{Introduction}	
 Heusler-based intermetallic alloys and compounds continue to be in the forefront of active research  due to their multifaceted electronic and magnetic properties. They have already been identified as material for the development of spintronics devices, magnetic actuators and switches~\cite{nanda,felser,ozdogan1,krenke}. So far, the  major attention has been given to 3$d$ transition metal-based Heuser compounds~\cite{snehashish}.  An early study of the rare-earth (RE) based full Heusler compounds was reported for Pd$_2$DySn and Pd$_2$HoSn, which were found to order antiferromagnetically  below 5 K~\cite{ishikawa,johnson,malik}. On the other hand, Pd$_2$RESn (RE = Tm, Lu, Y, Yb) were found to be superconducting~\cite{ishikawa}. Recently, rare-earth-bismuth-based full Heuslers and half Heuslers have attracted huge attention for their potential contribution in the field of spintronics. A number of REPdBi and REPtBi compounds were identified  as  topological  insulators or semimetals  with  nontrivial  band  topology ~\cite{lin,sawai}.  Most importantly, REPtBi and REPdBi are identified as Weyl semimetals showing anomalous Hall effect due to Berry curvature.~\cite{shekhar1,shekhar2}.

\par
A close relative to the REPdBi series is the  REPd$_2$Bi-type full Heusler compounds. However, unlike the half-Heuslers, REPd$_2$Bi compounds possess inversion symmetry. Most of them either order antiferromagnetically at low temperature (below 10 K) or show a nonmagnetic ground state~\cite{rpd2z,gofrykthesis}. The Dy, Ho and Er compounds in the series show additional signature of the first order phase transition in the resistivity versus temperature data around 100 - 150 K~\cite{rpd2z}. For the Ho-compound, the first order phase transition is found to be associated with a structural change as evident from the appearance of additional peaks in the temperature-dependent x-ray diffraction data~\cite{gofrykthesis}. To the best of our knowledge, there is no report of the magnetic and transport properties of  GdPd$_2$Bi in the literature. 

\par     
Gd has total angular momentum, $J = S =$ 7/2 due to the half-filled $4f$ shell, which imply that the magneto-crystalline anisotropy will be weaker. In the present work, we have performed a comprehensive study on  GdPd$_2$Bi through our magnetic, transport and structural investigations along with density functional theory (DFT) based  electronic structure calculations. Though there are several reports on the electronic structure of REPdBi and REPtBi compounds, the full Heusler based REPd$_2$Bi compounds are hardly investigated through {\it ab-initio} based electronic structure calculations. Our work identifies a martensitic type structural transition in the compound from a cubic L$2_1$ to orthorhmbic $Pmma$ structure below about 150 K. The compound shows anomalous Hall coefficient and unconventional negative magnetoresistance, which can be attributed to the presence of multi-band electronic structure.
	\begin{figure}[t]
		\centering
		\includegraphics[width = 8 cm]{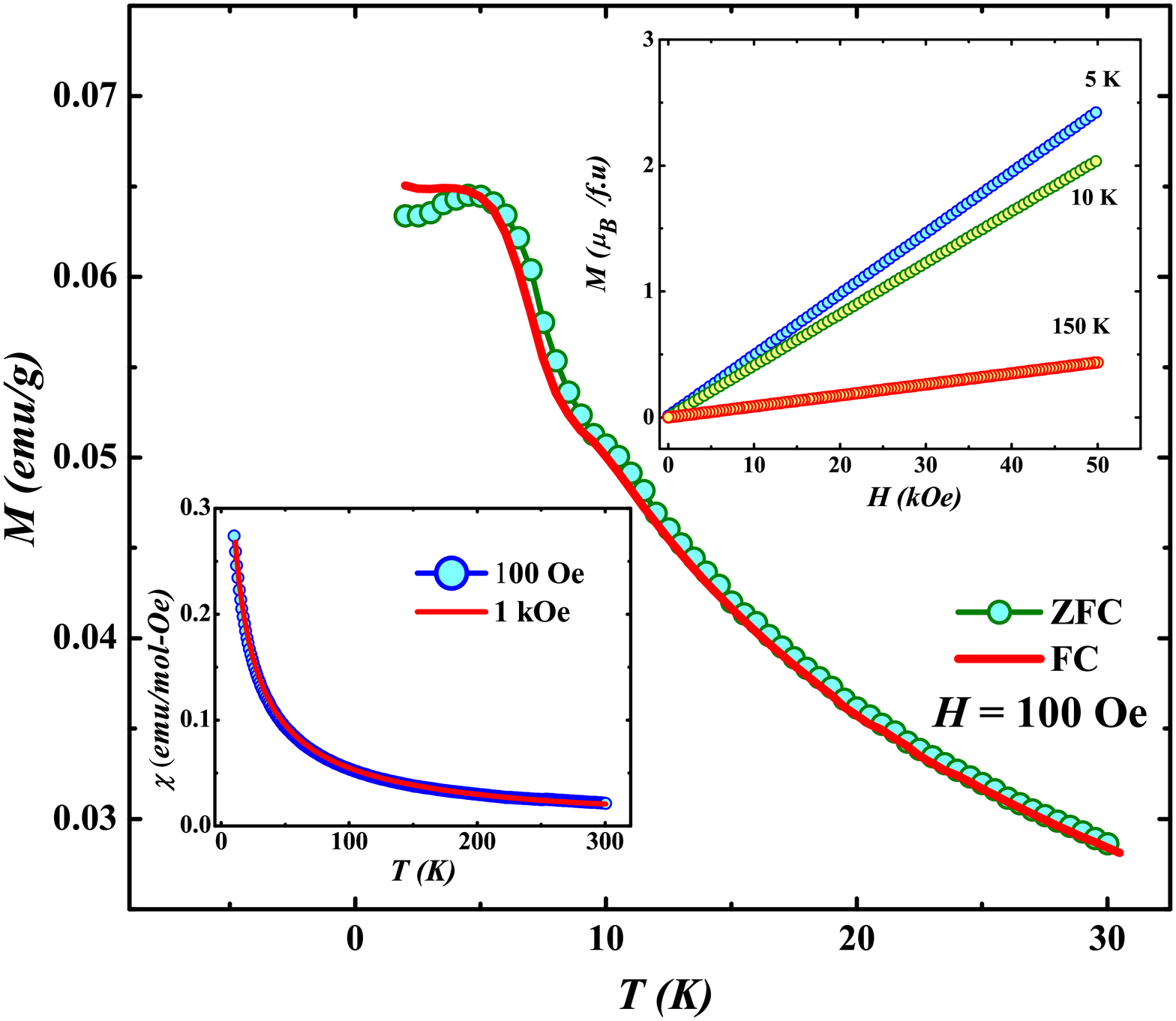}
		\vskip 2 cm
		\caption {$M$ vs $T$ curve measured at $H$ = 100 Oe both in ZFC and  FC protocols for GdPd$_2$Bi. Upper inset: isothermal magnetization data measured at different temperatures. Lower inset: magnetic susceptibility vs temperature data for $H$ = 100 Oe and 1 kOe.}
		\label{mag}
	\end{figure}
	\section{Experimental Details}
	Polycrystalline sample of GdPd$_2$Bi was prepared using conventional arc melting technique and the as cast ingot was used for further measurements. The phase purity and  the crystallographic structure were ensured by  powder x-ray diffraction (PXRD) followed by  Rietveld refinement, which  was performed using MAUD program~\cite{maud}. Rietveld refinement of the PXRD pattern obtained at room temperature confirms the formation of MnCu$_2$Al-type L2$_1$ structure (space group $Fm\bar{3}m$). The refined cubic lattice parameter is found to be 6.812(6) ~\AA.
	\par
	Magnetic measurements were carried out by using a commercial Quantum Design SQUID magnetometer (MPMS XL Ever Cool model). The resistivity ($\rho$) was measured by four probe method on a cryogen-free high magnetic field system (Cryogenic Ltd., U.K.) between 5-300 K. Thermopower measurement was performed in a lab based set up using differential technique between 10 K to 300 K. The sample was further investigated through high resolution temperature dependent x-ray diffraction (wavelength of the radiation being 0.749~\AA) using synchrotron facility at the Photon Factory, National Laboratory for High Energy Physics (KEK), Japan at various sample temperatures ranging from 15 to 300 K. Heat capacity ($C_P$) measurement was carried out using a Quantum Design Physical Properties Measurement System.
	
\section{Theoretical Techniques}
DFT-based electronic structure calculations were performed using the plane-wave basis set based on a pseudo-potential framework as incorporated in the Vienna $\textit{ab-initio}$ simulation package (VASP)~\cite{Kresse1993,Kresse1996}. The exchange-correlation functional was employed following the Perdew-Burke-Ernzerhof (PBE) prescription~\cite{Perdew1996}. The effect of the spin-orbit coupling (SOC) is introduced as a full-relativistic correction term to the Hamiltonian. For the plane-wave basis, a 350 eV cut-off was applied. To take care the missing onsite Coulomb, we have used $U_{eff}$ = 6 eV for the Gd, where  $U_{eff} = U-J_H$, $U$ is the onsite Coulomb interaction and $J_H$ is the Hund's coupling term~\cite{Anisimov1993,Dudarev1998}. The structural optimization was performed by relaxing the internal atomic positions toward the equilibrium until the Hellman-Feynman force becomes less than 0.001 eV~\AA$^{-1}$. A k-point mesh of 6 $\times$ 6 $\times$ 6 in the Brillouin zone (BZ) and the electronic convergence criteria were set to be at 10$^{-7}$ eV for self-consistent calculations.

\section{Experimental Results}
\subsection {Magnetization}
	GdPd$_2$Bi is found to order antiferromagnetically at $T_N$ = 7 K as shown in the main panel of fig~\ref{mag}. A weak bifurcation between field-cooled (FC) and zero-field-cooled (ZFC) magnetization ($M$) was observed below about 4 K. The dc magnetic susceptibility ($\chi = M/H$, $H$ is the applied magnetic field) between 20 to 300 K can be well-fitted with the Curie-Weiss law: $\chi(T) = C/(T-\theta_p)$, where $C$ is the Curie constant and $\theta_p$ is the paramagnetic Weiss temperature. From the good fitting of the $\chi$ vs $T$ [not shown here], we get the effective paramagnetic moment, $\mu_{eff}$  = 8.0 $\mu_{B}$/f.u and $\theta_P$ = $-$18 K. The  half-filled $4f$ shell of the free Gd$^{3+}$ has the effective moment  7.9 $\mu_B/f.u$, which is close to the observed value. The negative value of $\theta_P$ confirms the  antiferromagnetic (AFM) correlation in the system. The isothermal magnetization curves [upper inset of fig~\ref{mag}], measured up to $H$ = 70 kOe,  are found to be linear at $T$ = 5, 10 and 150 K. The linear isotherm even at 5 K supports the AFM nature of the magnetic ordering.
	
	\begin{figure*}[t]
		\centering
		\includegraphics[width = 12 cm]{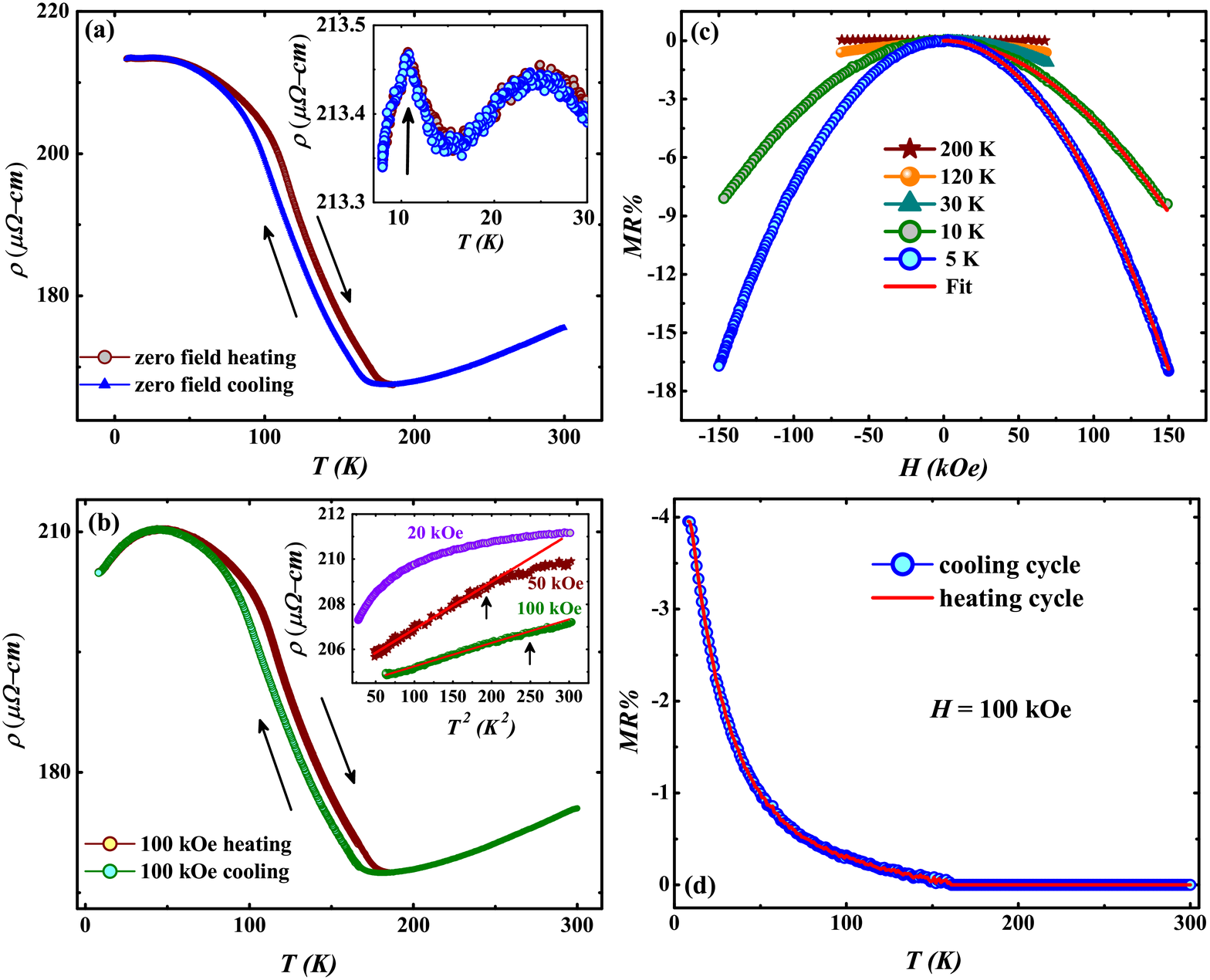}
		\caption {The electrical resistivity as a function of temperature at (a) $H$ = 0 kOe and (b) 100 kOe for GdPd$_2$Bi. Insets: (a) the enlarged view of the low-$T$ region of $\rho(T)$ measured at $H$ = 0 kOe and (b) the $T^2$ behaviour of the low-$T$ regime at different $H$. (c) the magnetoresistance as a function of applied magnetic field measured at different constant temperatures. The red line is the $H^2$ fit to the magnetoresistance data. (d) The magnetoresistance as a function of temperature measured at $H$ = 100 kOe.}
		\label{res}
	\end{figure*}
	\subsection{Electrical Resistivity}
	The most fascinating observation in the present work is obtained from the transport measurement. The $T$-variation of $\rho$ for GdPd$_2$Bi in presence of $H$ = 0 and 100 kOe is presented in figs.~\ref{res} (a) and (b) respectively. The sample shows conventional decrease in $\rho$ with temperature down to around $T$ = 180 K. On further cooling, $\rho$ follows an upturn with decreasing $T$ down to around 30 K and eventually it shows a saturating tendency. The signature of AFM ordering is also present in the zero-field low temperature data [see the inset of fig.~\ref{res} (a)] The region of upturn shows clear thermal hysteresis indicating the transition to be first order in nature. Similar upturn was previously observed in case of other members of the family, namely HoPd$_2$Bi and DyPd$_2$Bi~\cite{rpd2z}. Since we do not observe any feature between 100 - 200 K in the $M$ vs $T$ data, the observed transition is likely to be purely structural in nature. 
	
	\par
	The thermal hysteresis and the upturn survive even under an applied field of 100 kOe, and there is no significant change in the position of the upturn and the width of the thermal hysteresis. However, we observe a decrease in the value of $\rho$ under $H$ signifying negative magnetoresistance. Notably, a clear drop in $\rho$ is observed below about 40 K [as evident from fig~\ref{res}(b)], which indicates the  emergence of a magnetic field-induced metallicity in GdPd$_2$Bi.  The peak due to the AFM ordering in zero field [inset of fig.~\ref{res} (a)] vanishes for $H \geq$ 20 kOe. Inset of fig.~\ref{res}(b) shows the $\rho$ versus $T^2$ plot below $T$ = 20 K for different values of applied $H$. The system deviates completely from $T^2$ behaviour for $H <$ 20 kOe. It is evident that $\rho(T)$ follows a $T^2$ dependence for $H$ = 100 kOe below about 16 K, while similar $T^2$ variation is observed below 14 K for $H$ = 50 kOe. Such $T^2$ variation of $\rho$ is commonly attributed to electron-electron scattering in the Fermi liquid theory of metals. The coefficient of  the $T^2$  term is found to be $A$ =  1.3 $\times$ 10$^{-8}$ and 2.4 $\times$ 10$^{-8}$ $\Omega$-cm K$^{-2}$ at 100 and 50 kOe, respectively. The value of $A$ is about one order of magnitude higher than the conventional metals such as Cu~\cite{schatterjee1}.
	
	\par 
	The region of upturn in the $\rho(T)$ data around 180 K is associated with the negative temperature coefficient of resistivity ($d\rho(T)/dT <$ 0), and such behavior is generally observed in insulators, semiconductors or semimetals. However, the present rise can be ascribed to the structural transition, which is evident from our later analysis. It is interesting to note that similar upturn in resistivity is observed in case of several Ni-Mn-Z ($Z$ = In, Sn , Sb)-based shape memory alloys around the martensitic transition~\cite{brown2,chatterjee1,chatterjee2,chatterjee3}.  
	
	\par
To shed more light on the occurrence of field-induced metallicity, we have measured the field variation of $\rho$ up to $\pm$150 kOe at $T$ = 5 K and 10 K that eventually leads to a large negative magnetoresistance [$MR$ = $[\rho(H) - \rho(0)]/\rho(0)$]  of around $-$16\% and $-$9\% respectively. Hence, a conventional positive contribution from the Lorentz force can be ruled out for GdPd$_2$Bi. Interestingly, the negative MR in bulk GdPd$_2$Bi sample is `non-saturating' in nature and follows the form -$\frac{\Delta\rho}{\rho} \propto H^2$ [fig.~\ref{res}(c)] for $H$ as high as 150 kOe. In addition, it is interesting to note that the negative MR persists in the compound for $T$ as high as 120 K, which is much higher than the magnetic transition temperature ($T_N$ = 7 K), although the magnitude of MR decreases with increasing $T$ [fig.~\ref{res} (d)].
	
\par	
The MR can also be discussed in the framework of the semiclassical Kohler approach. In a single-band system with uniform scattering at all points on the Fermi surface, Kohler's scaling is often found to be valid and is expressed as
		\begin{equation}
			\frac{\Delta\rho}{\rho} = f \left( \frac{\mu_0H}{\rho} \right)
			\label{kohler}
		\end{equation}
where $f(x)$ is a temperature-independent implicit function~\cite{kohler}. $\Delta \rho/\rho$ versus $H/\rho$ measured at different temperatures should fall into a single curve. However, deviation from the general Kohler plot can be attributed to the reconstruction of Fermi surface for nesting and/or to the existence of multi-carrier electronic transport~\cite{ba8ge43,tmnic2,wte2}. Interestingly the title compound is found to violate the Kohler's rule (not shown here) which might be due to the existence of two type of carriers.
	
	\begin{figure}[t]
		\centering
		\includegraphics[width = 8 cm]{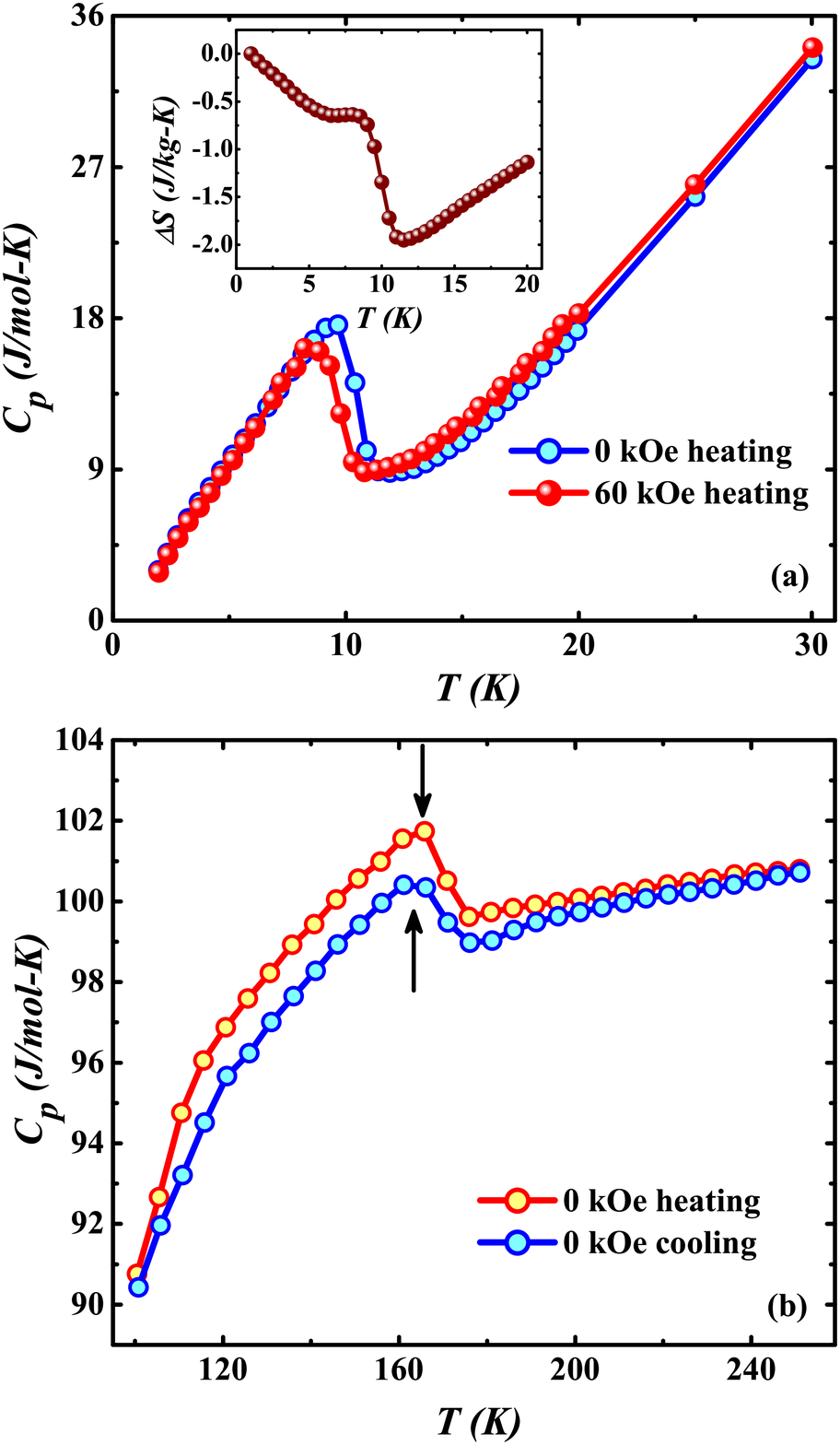}
		\caption {(a) Heat capacity ($C_p$) as a function of temperature in the range (a) 2 K-30 K and (b) 100 K-200 K measured under 0 kOe and 60 kOe for GdPd$_2$Bi. Inset of (a) shows $\Delta S$ versus $T$ plot.}
		\label{Cp}
	\end{figure}
	\subsection{Heat capacity}
	Fig.~\ref{Cp}(a) presents the low temperature $C_p(T)$ data measured in zero magnetic field as well as under $H$ = 60 kOe. The zero-field  data show a clear $\lambda$-like anomaly around 9.5 K, which matches well with the AFM transition observed in the magnetization data. Under the application of 60 kOe magnetic field, the transition does not get suppressed, albeit the peak slightly shifts towards lower $T$ (8.5 K). We have also calculated the change in entropy ($\Delta S$) due to the  the application of $H$ (magneto-caloric effect)~\cite{mce} around $T_N$ using the following relation.
	\begin{equation}
		\Delta S(T) = \int_0^T\left [{C_p(H)/T- C_p(0)/T}\right]dT.
		\label{magcal}
	\end{equation}
	
	$\Delta S(T)$ is found to be negative with a sharp anomaly at the magnetic transition temperature [inset of fig.~\ref{Cp}(a)].  $\Delta S$ attains a moderate value of 2 J/kg-K at $T_N$, which is connected to the field induced shift of $T_N$ to lower $T$.
	\par
	At high temperature, $C_p$ vs $T$ data shows a peak around $T_p$ = 162 K for both 0 kOe and 60 kOe measurements [fig.~\ref{Cp}(b)]. The peak in $C_p$ occurs around the temperature where the upturn in the  $\rho(T)$ data is observed. Such feature is associated with the structural phase transformation as mentioned in section F. The $C_p$ data around $T_p$ do not show any change due to the application of $H$.  Interestingly, the heating and the cooling data show thermal hysteresis around this structural transition, which mimics the similar observation in the $\rho(T)$ measurements. The $C_P$ value at 300 K is found to be pretty close  to the value obtained by using the Dulong-Petit method (= 3$pR$, where $R$ is the universal gas constant and $p$ (= 4 here) is the number of atoms per formula unit) for the sample ($\sim$ 100 J/mol-K).
	\begin{figure}[t]
		\centering
		\includegraphics[width = 7 cm]{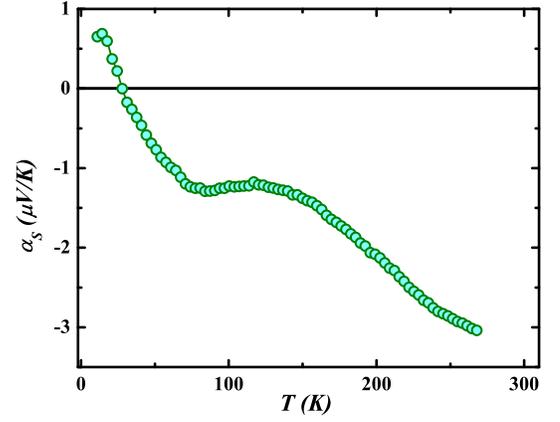}
		\caption {Seebeck coefficient as a function of temperature for GdPd$_2$Bi. }
		\label{thermo}
	\end{figure}
	\begin{figure}[t]
		\centering
		\includegraphics[width = 7 cm]{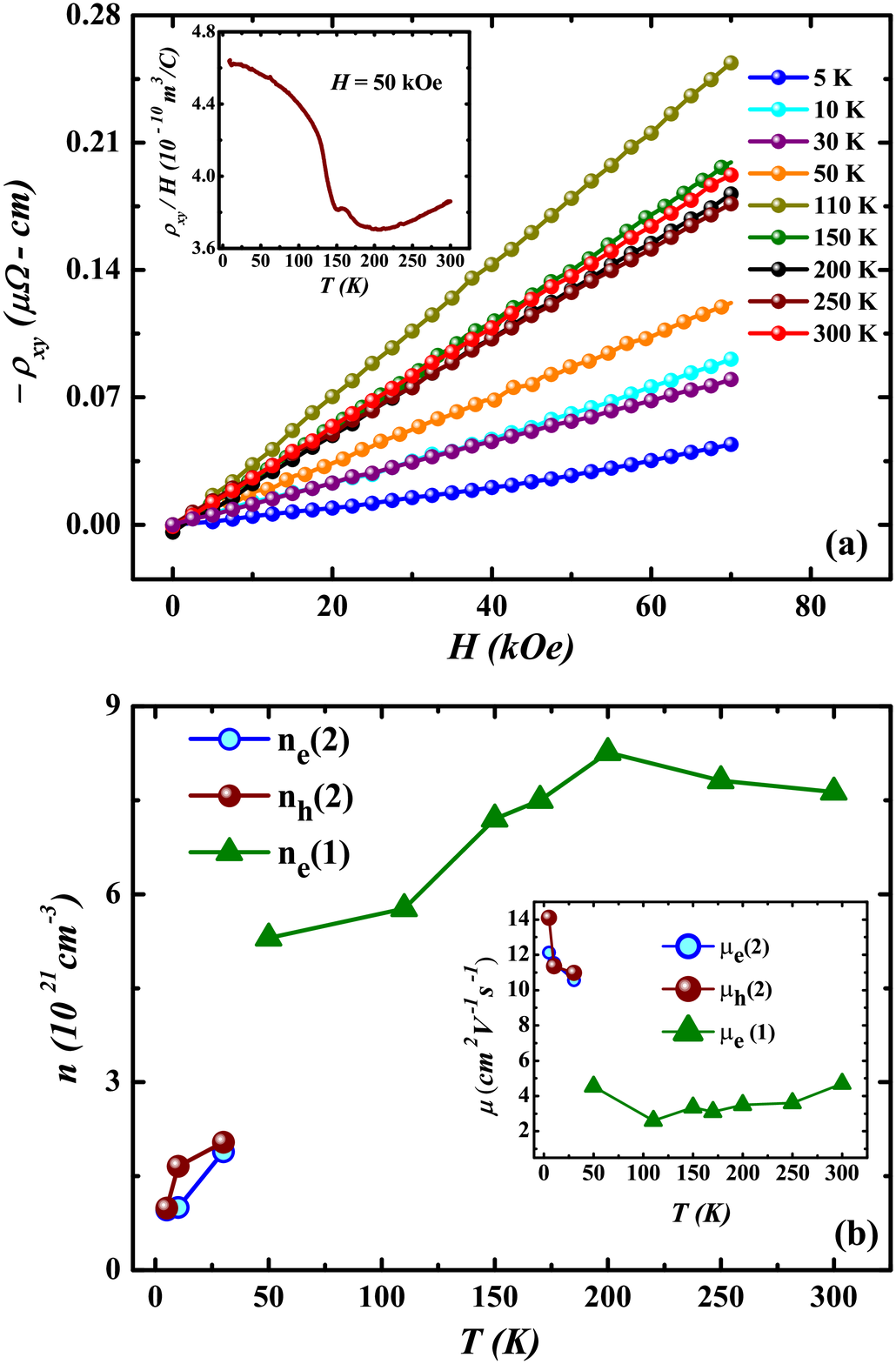}
		\caption { (a) $\rho_{xy}$ vs $H$ at different constant temperatures. Inset shows $T$ variation of $\rho_{xy}/H$ for $H$ = 50 kOe. (b) $T$-variation of carrier concentration. Inset: $T$ variation of Hall mobility for GdPd$_2$Bi. $n_{e,h}$(i) and $\mu_{e,h}$(i) denote carrier concentration and carrier mobility obtained from single carrier (i = 1) and two carrier (i = 2) model, respectively.}
		\label{hall}
	\end{figure}
\subsection{Thermopower} 
We have performed the thermopower measurement in terms of Seebeck coefficient ($\alpha_S$)~\cite{thermo}  in zero field as shown in  fig.~\ref{thermo}. GdPd$_2$Bi shows negative value of $\alpha_S$ between 25 - 270 K attaining a maximum  value with magnitude  $\sim$ 3 $\mu$V/K. $\alpha_S(T)$ shows a hump-like feature around 140 K, which matches well with the feature observed in the $\rho(T)$ data. Interestingly, $\alpha_S$ changes its sign and turns positive below 25 K and attains a value of $\sim$ 1 $\mu$V/K at 12 K, which may be due to the  presence of both electrons and holes in the system. In a two-carrier system the thermopower is often expressed as
	\begin{equation}
		\alpha_S = \frac{{\sigma_h\alpha_{S,h}} + {\sigma_e\alpha_{S,e}}}{{\sigma_h + \sigma_e}}
		\label{seebeck}
	\end{equation}
	where $\sigma$ is the electrical conductivity and the subscripts $h$ and $e$ denotes the hole and electron contributions, respectively. Thus, eqn~\ref{seebeck} makes it clear that, $\alpha_S(T)$ might exhibit change in sign with temperature in the case of a two carrier system~\cite{ba8ge43}. Notably, we find that  GdPd$_2$Bi is the only compound in  the REPd$_2$Bi series which predominantly shows negative value  of  $\alpha_S$~\cite{rpd2z}. On the other hand, REPd$_2$Sb compounds show negative $\alpha_S$ value but they do not undergo any structural transition (lack of thermal hysteresis)~\cite{rpd2z,erpd2sb}. Bi-based compounds are often found to be potential candidates for thermoelectric applications~\cite{rpdbi}. But in case of GdPd$_2$Bi, the power factor (PF) is rather low ($\sim$ 25 $\mu$W m$^{-1}$K$^{-2}$).

\subsection{Hall measurements}
The Hall resistivity ($\rho_{xy}$) for GdPd$_2$Bi was determined as the antisymmetric part of the measured transverse voltage, $\rho_{xy}$ = t[$V_{xy}$(+$H$) - $V_{xy}$(-$H$)]/2$I$, where $t$ is the thickness of the sample, $I$ is the applied current and $V_{xy}$ is the transverse voltage generated. Variation of $\rho_{xy}$ with $H$ measured at different constant temperatures up to $H$ = 70 kOe is depicted in fig.~\ref{hall}(a)~\cite{hall1,hall2}. The negative value of $\rho_{xy}$ at all measured temperatures indicates  electron to be the majority carrier in the system. As temperature is lowered from room temperature, $\rho_{xy}$ vs $H$ show linear variation  above  about 30 K. However, on further lowering of $T$, $\rho_{xy}$ turns non-linear for $T \leq$ 30 K suggesting that a simple picture of the single-carrier model for the conventional Hall effect  is inadequate for the sample, at least below 30 K. 
\par
We have depicted the $T$-variation of $\rho_{xy}$/$H$ in the inset of fig.~\ref{hall}(a), and it is strongly $T$-dependent along with a clear anomaly around the temperature of thermal hysteresis observed in the $\rho(T)$ data.

\par
\begin{figure}
	\centering
	\includegraphics[width = 6 cm]{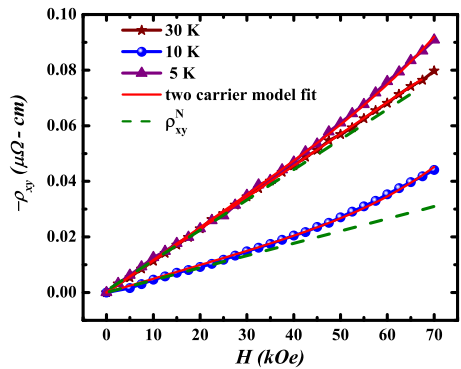}
	\caption {$\rho_{xy}$ vs $H$ for $T$ = 5, 10 and 30 K along with the fit of two carrier model (eqn~\ref{twoband}). $\rho^N_{xy}$ denotes the linear Normal Hall contribution.}
	\label{hallfitting}
\end{figure}

As mentioned earlier, the Hall resistivity at low-$T$ ($T$ = 5 K, 10 K and 30 K) shows a non-linear nature and follows a concave upward shape as can be seen from the fig~\ref{hallfitting}. The non-linearity at low-$T$ may arise either due to the presence of more than one charge carrier or due to the anomalous Hall contribution~\cite{euagas}. Since the sample shows a perfect linear variation of $M$ with $H$ [see upper inset of fig.~\ref{mag}], the nonlinearity of the $\rho_{xy}(H)$ cannot be simply attributed to the anomalous Hall effect (AHE) of magnetic origin. Notably, $\alpha_S$ turns from negative to positive below about 25 K. Such behavior may be due to the  presence of both electrons and holes in the system. Therefore, a multiband model consisting of holes and electrons seems logical to describe the nonlinearity in $\rho_{xy}$.       

We invoked  the following two carrier model to fit the Hall resistivity~\cite{model3}: 
\begin{equation}
	\rho_{xy}(H) = \frac{H}{e}\left [\frac{(n_h\mu_h^2 - n_e\mu_e^2)+(n_h - n_e)(\mu_h^2\mu_e^2)H^2}{(n_h\mu_h + n_e\mu_e)^2+{(n_h - n_e)^2\mu_h^2\mu_e^2}H^2} \right]
	\label{twoband}
\end{equation}
Here $n_h$, $n_e$, $\mu_h$ and $\mu_e$ are the carrier density and mobility of hole(h) and electron(e) respectively. As can be seen from fig.~\ref{hallfitting}, $\rho_{xy}$  for $T$ = 5, 10 and 30 K can be fitted with eqn.~\ref{twoband} suggesting that the two carrier model can successfully describe the experimentally measured data below about 30 K. However, the linear nature of $\rho_{xy}$ versus $H$ above 30 K prompted us to express the Hall resistivity data via an effective single-carrier model, where the normal Hall-coefficient ($R_H$ = $\rho_{xy}(H)/H$) is calculated from the slope of high-field (40 kOe $< H <$ 70 kOe) data. Hence, $\rho_{xy}$  is fitted with two different models for $T <$ 30 K regime and $T >$ 30 K temperature range, where two-carrier and single-carrier models respectively, gives best fit of the experimental data~\cite{yvsn6,srcdbi2}. 
\par
The temperature dependence of the carrier concentration ($n_e$ and $n_h$) and Hall mobility ($\mu_e$ and $\mu_h$) are shown in fig.~\ref{hall}(b). The extracted value of electron carrier concentration increases from 9.6 $\times$ 10$^{20}$ cm$^{-3}$ at 5 K (derived from two carrier model) to 7 $\times$ 10$^{21}$ cm$^{-3}$ (obtained from single carrier model, $n_e = \frac{1}{R_H e}$) at room temperature for GdPd$_2$Bi which is in line with the typical of semimetallic materials~\cite{aliev,semi-hall}. The Hall mobility ($\mu$) is also deduced from the linear fitting of the $\rho_{xy}$ data using the equation: $\mu$ = $R_H$/$\rho$($H$ = 0). $\mu$ decreases with increase in $T$ and then saturates at higher temperatures above about 110 K [see inset of fig~\ref{hall}(b)]. With lowering of $T$, $n_e$ increases slightly down to 200 K and then decreases sharply, around the temperature range where the thermal hysteresis is observed. Thereafter, it continues to  decreases down to 5 K.
\subsection{Temperature evolution of structural parameters}
\begin{figure*}[t]
	\centering
	\includegraphics[width = 14 cm]{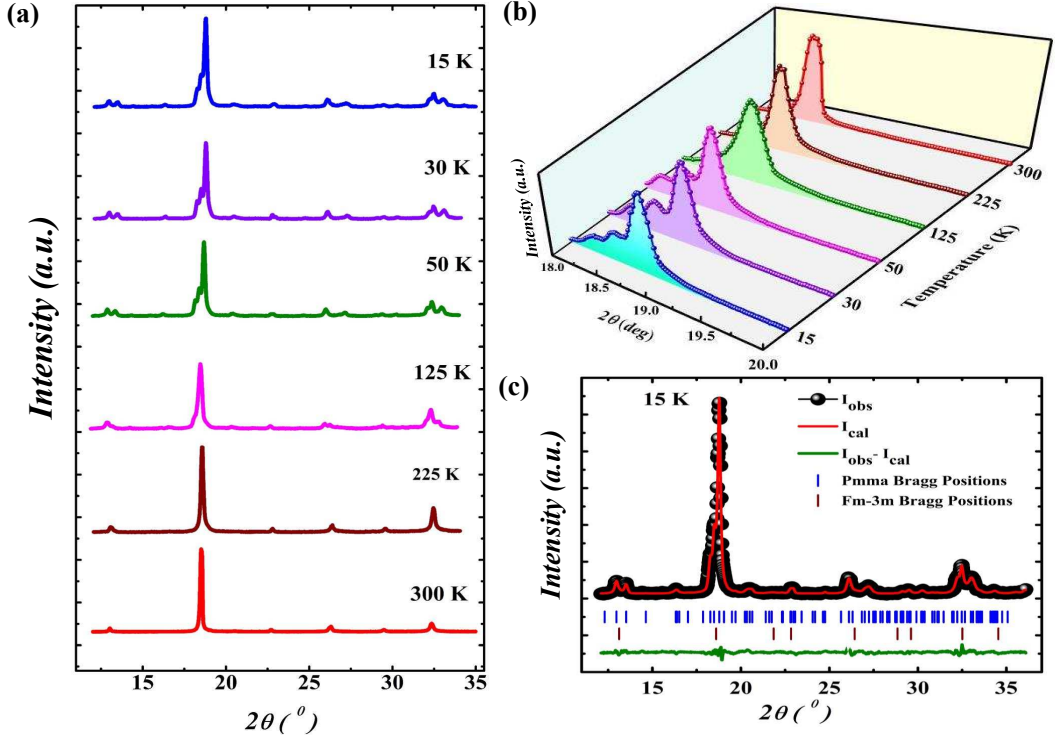}
	\caption {(a) X-ray diffraction patterns for GdPd$_2$Bi at various temperatures. (b) Magnified 3-D plot for the most intense peak for various temperatures. (c) X-ray diffraction pattern recorded at 15 K; observed(black circles), calculated (continuous red line), difference (continuous green line) intensities and allowed Bragg positions (blue ticks and brown ticks) are marked in the figure.}
	\label{xrd}
\end{figure*}

 Synchrotron diffraction performed at different constant temperatures are plotted in fig.~\ref{xrd}(a). It shows [see figs.~\ref{xrd}(a) and (b)] the change in the pattern as $T$ is lowered from 300 K to 15 K. To highlight the change, the temperature evolution of the most intense (220) peak of the cubic phase  is shown in fig.~\ref{xrd}(b). Splitting of the main peak with the decrease in $T$ is consistent with the loss of cubic symmetry, and the structure presumably attains a lower symmetry. Earlier Gofryk {\it et. al.} proposed the existence of a  structural transition in the REPd$_2$Bi series of compounds~\cite{rpd2z}, and the low-$T$ crystal structure was reported to be orthorhombic ($Pnma$)~\cite{gofrykthesis}. 
 
 \par
\begin{figure}[t]
\centering
\includegraphics[width = 9 cm]{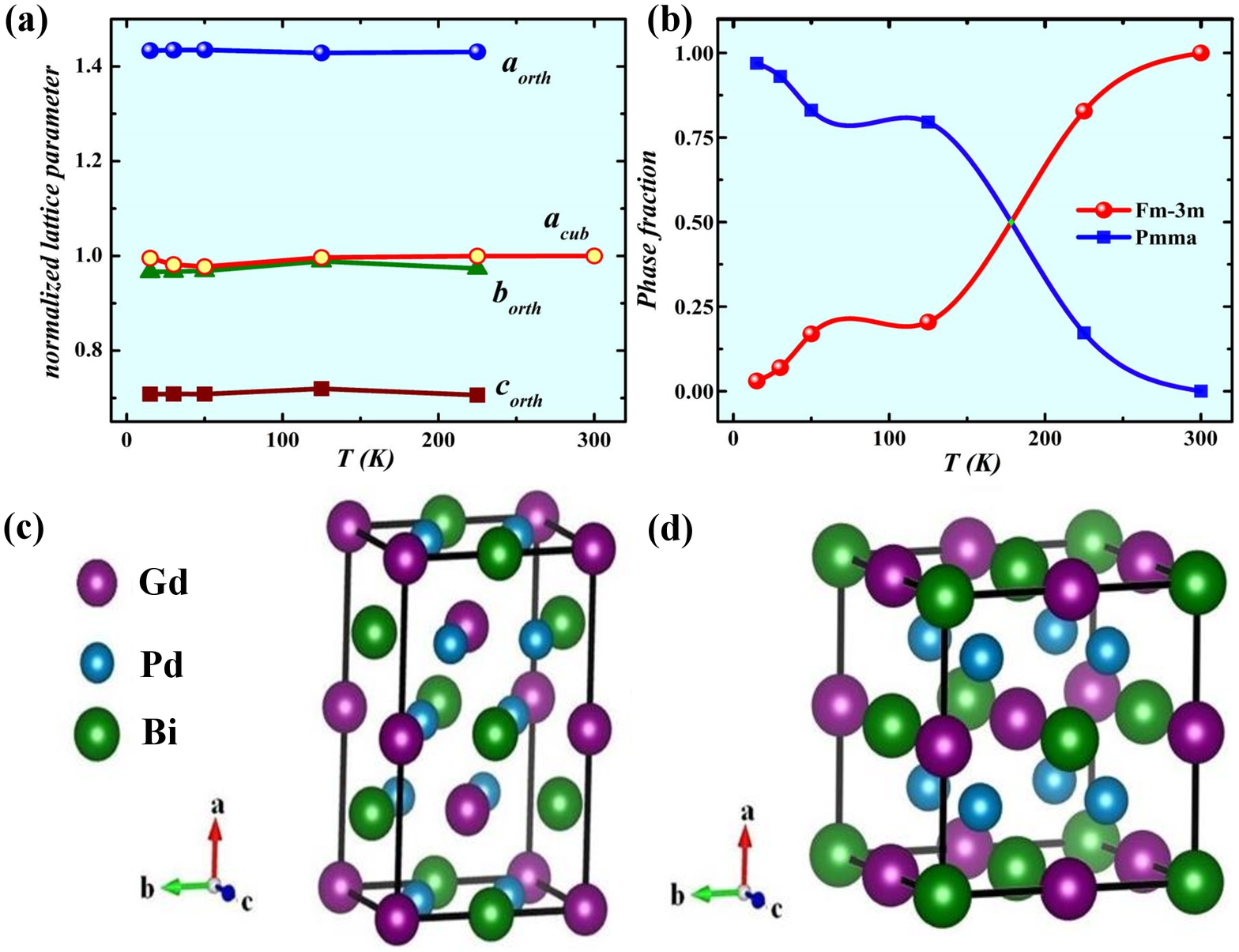}
\caption {(a) The thermal variation of lattice parameters ($a_{orth}$, $b_{orth}$, $c_{orth}$ and $a_{cub}$), (b) phase fraction $\%$ with temperature. (c) and (d) the crystallographic image of low-$T$ (Pmma) and high-$T$ (Fm$\bar{3}$m) phase respectively.}
\label{latpar}
\end{figure}

 We have carefully analyzed the data using Rietveld refinement technique and the fitted data for the representative temperature 15 K are shown in fig.~\ref{xrd}(c). We used \texttt{ISODISTORT}~\cite{isodistort1,isodistort2} software to identify all the possible list of non-cubic, non-isomorphic subgroups of the  parent cubic structure. We find that the orthorhombic $Pmma$  generates all the experimentally observed reflections correctly and converge well with the  data. In the intermediate temperatures, both cubic  and orthorhombic structures co-exist and we have considered both the phases to refine the  diffraction data. The thermal evolution of the lattice parameters is plotted (with respect to the cubic lattice parameter, $a_{cub}$) in fig.~\ref{latpar}(a). The refined lattice parameters for the cubic (at 300 K) and orthorhombic (at 15 K) phases are found to be $a_{cub}$ = 6.812(6) \AA~ and $a_{orth}$ = 9.766(1) \AA~, $b_{orth}$ = 6.582(5) \AA~ and $c_{orth}$ = 4.824(1) \AA~ respectively. In case of a cubic to orthorhombic structural phase transition, the lattice parameters for the low-temperature (LT) orthorhombic unit cell is associated to the high-temperature (HT) cubic structure by the relation: $a_{orth}$ = $\sqrt{2}a_{cub}$, $b_{orth}$ = $a_{cub}$ and $c_{orth}$ = $\sqrt{2}a_{cub}$~\cite{brown}. The $c/a$ ratio turns out be close to 0.75 indicating considerable distortion of the cubic cell as temperature is lowered. The cell volume around the thermal hysteresis regime changes  around 2\% between the low- and high-temperature phases which is similar to the cell volume change of the ferromagnetic shape memory alloy, Ni$_2$MnGa~\cite{brown}. The atomic positions for the cubic and orthorhombic structures are given in table~\ref{lat_par}. These values are similar to those of other Heusler-based shape memory alloys, such as Co$_2$NbSn~\cite{co2nbsn} and Ni$_{2.04}$Mn$_{1.4}$Sn$_{0.56}$~\cite{nimnsn}, undergoing martensitic type phase transition.
\par
For the 15 K data, there is a finite percentage of cubic phase present suggesting the co-existence of both the phases even at low-$T$ region. The variation of phase fraction consisting of the orthorhombic phase and cubic phase with temperature is presented in fig.~\ref{latpar}(b). Increase in the fraction of orthorhombic phase along with the decrease of cubic phase with decrease in temperature can be clearly observed. This observation supports the occurrence of phase transformation, as also evident from the thermal hysteresis in the $\rho$ vs $T$ data along with the anomaly in the high temperature $C_P$ data. 
\begin{table*}
	\caption{Crystallographic parameter of the sample as obtained from the refinement of the PXRD data.}
	\centering 
	\begin{tabular}{c c c}
		\Xhline{3\arrayrulewidth}
		Temperature (K) &300&  15      \\
		Structure/space group &L2$_1$ cubic $F$m$\bar{3}$m& Orthorhombic $Pmma$   \\
		Cell parameters &$a_{cub}$ = 6.812(6) \AA& ~~~$a_{orth}$ = 9.766(1) $b_{orth}$ = 6.582(5) $c_{orth}$ = 4.824(1) \AA~    \\
		Cell volume (\AA$^3$) &316.24(1)& 309.63(5)   \\
		Standard deviation ($\sigma_D$) &1.77&1.56     \\
		\Xhline{3\arrayrulewidth} \\
		Atoms&~~Site~~~~~x~~~~y~~~~z~~~~B$_{iso}$&~~~Site~~~~x~~~~~~~y~~~~~~z~~~~~~~~~B$_{iso}$ \\
		\Xhline{3\arrayrulewidth} \\
		{\footnotesize Gd} &~~~~~{\footnotesize 4b}~~~~~$\frac{1}{2}$~~~$\frac{1}{2}$~~~~$\frac{1}{2}$~~~{\footnotesize 0.5(3)} ~&~~~~{\footnotesize 2a}~~~~$\hspace{0.15 cm}${\footnotesize 0}~~~~~$\hspace{0.2 cm}${\footnotesize 0}~~~~~~$\hspace{0.15 cm}${\footnotesize 0}~~~~~~~{\footnotesize 1.1(2)}   \\
		~~~~~&~~~~~~~~~~~~~~&$\hspace{0.45 cm}${\footnotesize 2f}$\hspace{0.55 cm}$$\frac{1}{4}$~~~~$\hspace{0.25 cm}$$\frac{1}{2}$$\hspace{0.45 cm}${\footnotesize 0.509(1)}$\hspace{0.2 cm}${\footnotesize 0.9(1)}   \\ \\
		{\footnotesize Pd} &~~~~{\footnotesize 8c}~~~~~$\frac{1}{4}$~~~$\frac{1}{4}$~~~~$\frac{1}{4}$~~~{\footnotesize 0.4(4)} &~~~~{\footnotesize 4h}~~~$\hspace{0.25 cm}${\footnotesize 0}~~~{\footnotesize 0.266(5)}~~~$\frac{1}{2}$$\hspace{0.55 cm}$~~{\footnotesize 0.7(1)}  \\
		~~~~~&~~~~~~~~~~~~~~&$\hspace{0.65 cm}${\footnotesize 4k}~~~~~$\frac{1}{4}$~~~{\footnotesize 0.265(2)}~~{\footnotesize 0.031(9)}~{\footnotesize 0.45(9)}  \\ \\
		{\footnotesize Bi} &~~~~{\footnotesize 4a}~~~~~{\footnotesize 0}~~~{\footnotesize 0}~~~~~{\footnotesize 0}~~~{\footnotesize 1.1(9)}~~&$\hspace{0.35 cm}${\footnotesize 2b}$\hspace{0.5 cm}${\footnotesize 0}$\hspace{0.6 cm}$$\frac{1}{2}$$\hspace{0.9 cm}${\footnotesize 0} $\hspace{0.7 cm}${\footnotesize 1.2(1)}  \\ 	
		~~~~~&~~~~~~~~~~~~~~&$\hspace{0.4 cm}${\footnotesize 2e}~~~~~$\frac{1}{4}$~~~~~{\footnotesize 0}~~~~{\footnotesize 0.503(3)}~~~~{\footnotesize 1.1(5)}   \\ \\ \Xhline{3\arrayrulewidth}
	\end{tabular}
	\label{lat_par}
\end{table*}

\section{Theoretical Results}

\subsection{Density of states}
\begin{figure}
\centering
\includegraphics[width = 7 cm]{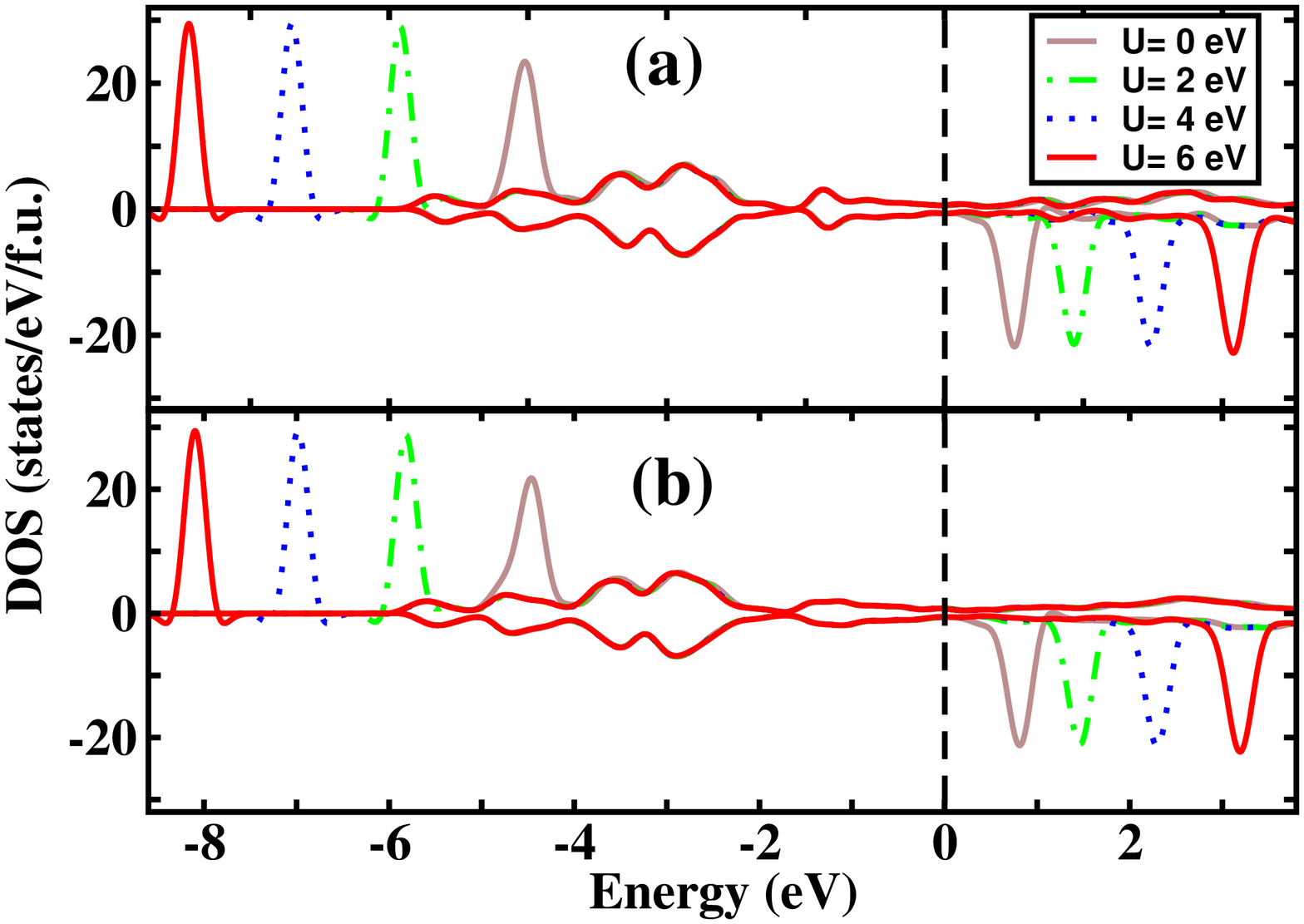}
\caption {The calculated spin-polarized DOS for (a) high temperature and (b) low temperature, for different $U_{eff}$ values. Two panels represent majority and minority spin channels for each DOS. The Fermi level is set at zero in the energy scale.}
\label{DFT-Fig1}
\end{figure}

The first principles electronic structure calculations have been done for both LT and HT phases of GdPd$_2$Bi. The calculated spin polarized GGA+$U_{eff}$  density of states (DOS) are shown in fig.~\ref{DFT-Fig1}(a) and fig.~\ref{DFT-Fig1}(b) for high and low temperature phases of GdPd$_2$Bi respectively. From the DOS, it is clear that both the high and low temperature phases are metallic in nature with small but finite DOS contributions at the Fermi energy in the both spin channels. The calculated spin magnetic moments at the position of Gd ions/site is $\approx$ 7.10 $\mu_B$/site for both high temperature cubic and low temperature orthorhombic structures. Whereas, the induced magnetic moment at the Pd and Bi is very small around -0.03$\mu_B$. Point to be noted that, the induced moment at the Pd and Bi sites are of opposite sign to that of the sign of the Gd magnetic moment. The electronic structures of both low and high temperature phases have been verified by varying the onsite Coulomb correlations i.e. $U_{eff}$ from 0 to 6 eV at the rare-earth element Gd site. As the onsite coulomb correlation increases, the Gd-$4f$ states are moving apart as shown clearly in fig.~\ref{DFT-Fig1}, without opening any gap  at the Fermi energy, confirming the robustness of the metallic nature of the electronic structure.

\subsection{Magnetic configuration}

\begin{figure}
\centering
\includegraphics[width = 6cm]{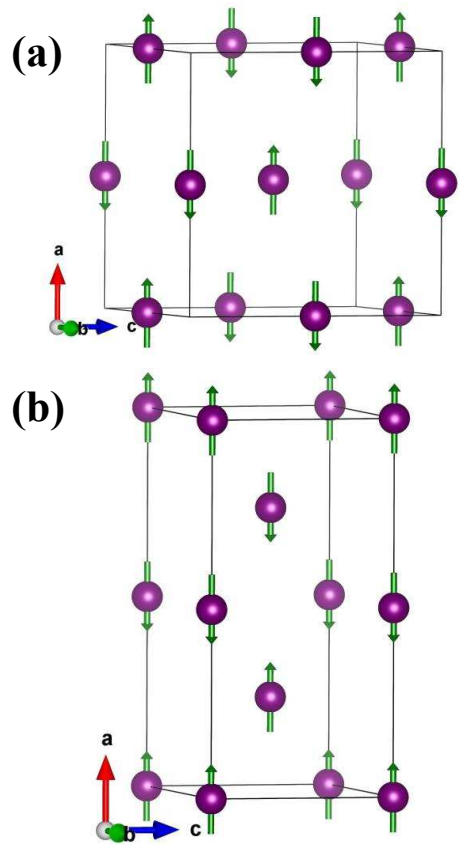}
\caption{The pictorial representation of the lowest energy antiferromagnetic spin configurations  for (a) 
HT and (b) LT of GdPd$_2$Bi in the Gd only sublattice structure.}
\label{DFT-Fig2}
\end{figure}

Experimentally, it has been found that the antiferromagnetic correlation for both high and low temperature. Therefore, it is pertinent to calculate the total energies of different spin configurations for both HT and LT structures. The calculated energetics are shown in table II. The results show that in both the structures the AFM configuration is energetically lower than the ferromagnetic configuration by 3.77 (2.95) meV/f.u. for HT (LT) structures (as mentioned in table 2). We have calculated various other spin configurations as well. However, the presented spin configurations in fig.~\ref{DFT-Fig2} are energetically lower than the other antiferromagnetic configurations. In the HT cubic structure, the Gd only sublattice as shown in the fig.~\ref{DFT-Fig2}(a) forms an regular tetrahedral unit with $\uparrow$ $\downarrow$ $\uparrow$ $\downarrow$ spin arrangement. On the contrary, in the case of LT orthorhombic structure, due to reduction of the symmetry, the Gd atoms form an isosceles triangle with antiferromagnetic arrangement. As a whole, the spins on the Gd atoms are arranged parallel in  the $\textit{bc}$ plane and coupled anti parallel along the crystallographic $\textit{a}$ direction.

\begin{table}
\begin{center}
\caption{Energetics of the different spin configurations of the HT and LT phases of GdPd$_2$Bi}
\begin{tabular}{|c|c|c|c|c|c|c|}
\hline
Phase  & Configuration & Gd1 & Gd2 & Gd3 & Gd4 & $\Delta$E (meV) \\
\hline
HT  & FM            & $\uparrow$ & $\uparrow$ & $\uparrow$ & $\uparrow$ & 3.77 \\
    & AFM          & $\uparrow$ & $\downarrow$ & $\uparrow$ & $\downarrow$ & 0.0 \\
\hline
LT  & FM            & $\uparrow$ & $\uparrow$ & $\uparrow$ & $\uparrow$ & 2.95 \\
    & AFM          & $\uparrow$ & $\downarrow$ & $\uparrow$ & $\downarrow$ & 0.0 \\
\hline                   
\end{tabular}
\end{center}
\label{spin1}
\end{table}      

\subsection{Band structure}
\begin{figure}
\centering
\includegraphics[width = 8 cm]{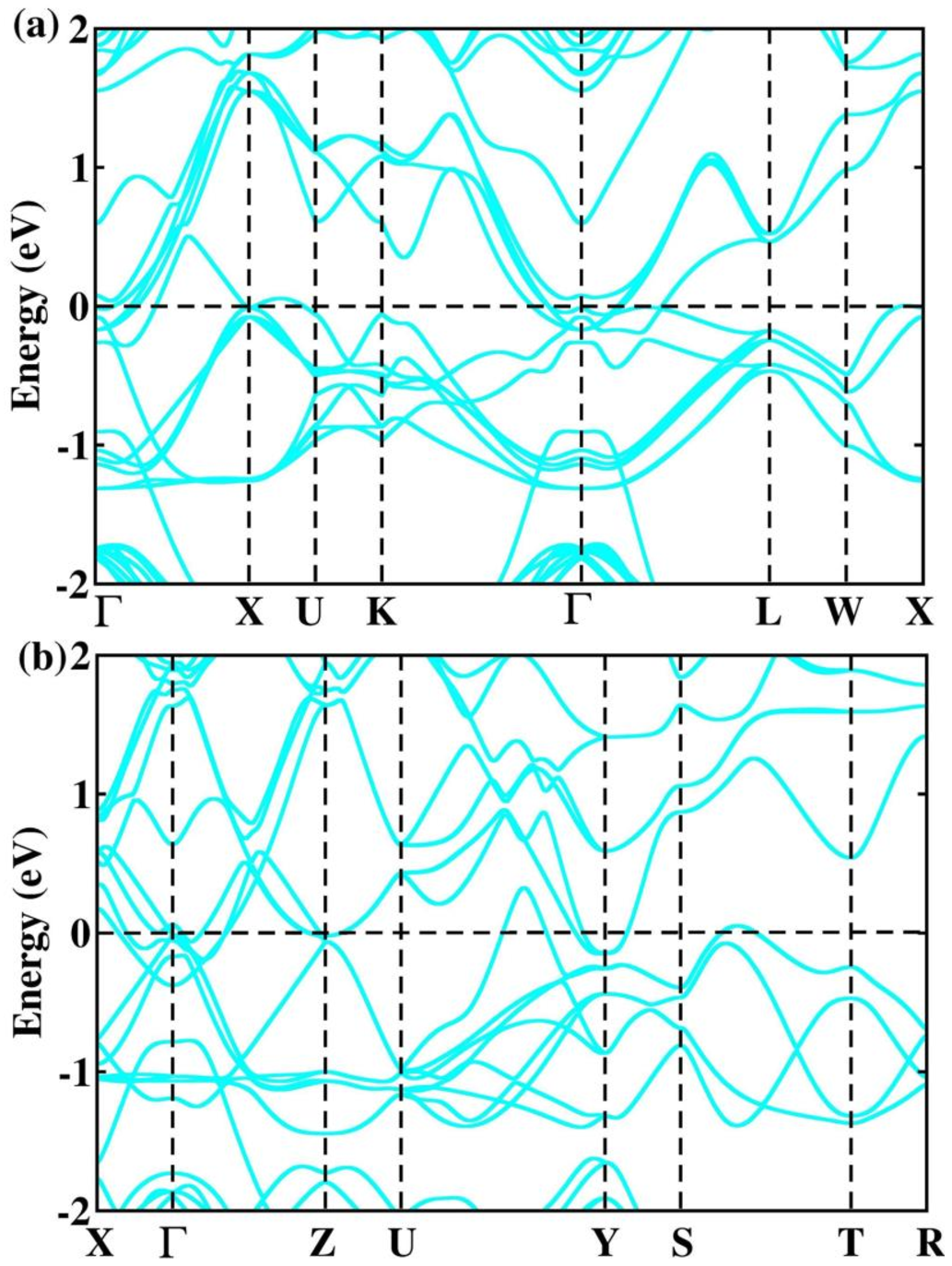}
\caption {The calculated spin-polarized GGA+$U$+SOC band structures along the high symmetry direction in the Brillouin zone (BZ) for (a) HT and (b) LT of GdPd$_2$Bi, respectively. The Fermi level is set at zero in the energy scale. }
\label{DFT-Fig3}
\end{figure}
The robustness of the metallicity and the electronic structure has been also analysed in the presence of the spin-orbit coupling for both HT and LT phases. The fig.~\ref{DFT-Fig3} shows the  band structure in presence of the Coulomb correlation $U$ and the SOC in the AFM ground state spin configurations. From the band structure it is very clear that even in the presence of the SOC and AFM correlation, the metallicity is preserved. Moreover, careful investigation reveals that for both the HT (fig.~\ref{DFT-Fig3}(a)) and LT (fig.~\ref{DFT-Fig3}(b)) configurations, the valence and conduction bands cross the Fermi energy. For example, in the HT cubic phase, the valence bands cross the Fermi energy in the $\textit{X-U}$ direction of the BZ, whereas the conduction bands in multiple region such as $\textit{K-$\Gamma$-L}$ and 
$\textit{$\Gamma$-X}$ in the BZ. Similarly, for the LT phases, the conduction bands cross the $E_F$ about the $\textit{$\Gamma$}$, $\textit{Z}$ and $\textit{Y}$ points in the BZ. However, the valence bands cross the $E_F$ along the $\textit{X-$\Gamma$-Z}$, $\textit{U-Y}$ and $\textit{S-T}$ directions in the BZ. The bands that cross the Fermi energy dominantly contribute from the Pd-$\textit{4d}$ and 
Bi-$\textit{5p}$ orbitals. The effect of SOC on the magnetic moment is minimal, except that there is small orbital magnetic moment at the Gd site of the order of 0.02 $\mu_B$.

\subsection{Fermi surface}
\begin{figure}
\centering
\includegraphics[width = 8.5 cm]{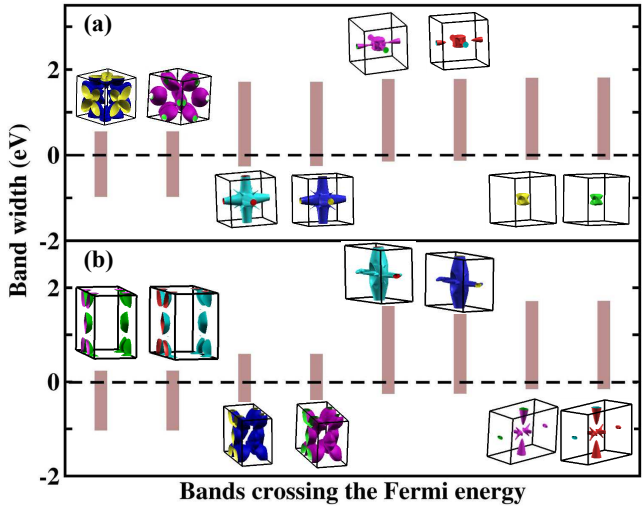}
\caption {The calculated GGA+$U$+SOC band widths for (a) HT and (b) LT of GdPd$_2$Bi are indicated by vertical brown bars which crosses the Fermi energy. The corresponding Fermi surfaces are shown in the insets.}
\label{DFT-Fig4}
\end{figure}

We have plotted the band width crossing the $E_F$, and the corresponding Fermi surfaces in presence of the SOC for both HT and LT phases in fig.~\ref{DFT-Fig4}(a) and (b), respectively. From the band widths, it is clear that both electron pockets and hole pockets are present in both HT and LT phases of the title compound. However, careful investigation of the fig.~\ref{DFT-Fig3} and ~\ref{DFT-Fig4}, confirms that the hole pocket is more dominant in the LT phase than the HT phase. In fig.~\ref{DFT-Fig4}(a), the bands crossing the Fermi energy and forming electron pockets are predominant in nature. The corresponding Fermi surfaces are also shown in the inset of fig.~\ref{DFT-Fig4}. From  fig.~\ref{DFT-Fig4}(a) insets, it is clear that the electron pocket Fermi surfaces are mainly three types, whereas- the hole pocket Fermi surfaces are of one type, with complicated nesting in the BZ. The electron pockets are mostly located around the $\Gamma$ point, and the hole pockets are at the edge of the BZ near the high symmetry $U$ and $X$ points.  Interestingly, although there are some hole pockets in the Fermi surface, the dominant carriers are still the in electron pockets in the BZ. The LT case is not exactly same as that of the HT in the context of the carrier pockets in the BZ. In fig.~\ref{DFT-Fig4}(b) insets, the Fermi surfaces of the electron and hole pockets are shown for the LT phase, which has few distinct difference compared to that of the HT phase. In the LT phase, the contribution of the electron pocket Fermi surfaces has decreased, whereas- the contribution of the hole pocket Fermi surfaces has increased, compared to that of the HT phase. In fig.~\ref{DFT-Fig4}(b), the hole pockets form two different kind of the Fermi surfaces located around the $\Gamma$ point and spans larger areas in the BZ. Therefore, from the Fermi surface plots, it is clear that in both HT and LT phases, electrons and holes contribute in the carrier type, however, in the LT phase, the hole contributions have increased compared to that of the HT phase.
\section{Discussion}
The structural transition from cubic to low-symmetry phase is already reported for REPd$_2$Bi (RE = Dy and Ho)~\cite{rpd2z}. However, Gd with orbital angular momentum $L$ = 0 at the ground state, is electronically different from the other rare-earth ions. Nevertheless, the observation of structural phase transition indicates that orbital moment has little effect on the phenomenon.
\par 
GdPd$_2$Bi depicts  thermal hysteresis in the electrical transport measurement around the structural transition. At high temperatures ($>$ 200 K), the $\rho(T)$ is metallic followed by an upturn below 200 K, and this region of upturn shows thermal hysteresis indicating a first order phase transition. Evidently, this upturn manifests the structural transition observed in our PXRD data [see fig~\ref{latpar}].  At lower temperatures below 30 K, the resistivity shows a saturating tendency, which eventually turns metallic under the application of $H$. Such sharp decrease in $\rho$ under $H$ in the low-$T$ region has also been reported in another half-Heusler compound HoNiSb, which is predominantly semiconducting in nature. In case of HoNiSb, such decrease in $\rho(T)$ in the low temperature, has been attributed to the reduction of spin-disorder scattering due to the alignment of moments under the application of $H$, and the reduction of the gap occurring due to the splitting of the up and down spin subbands~\cite{karla}. It is interesting to note that the $\rho(T)$ has remarkable similarities with several transition metal-based shape memory alloys undergoing martensitic phase transition (MPT), where sharp and hysteretic rise in $\rho$ between two metallic phases is  observed. The common examples are Ni-Co-Al, Ni-Mn-Sn, Ni-Mn-In etc~\cite{chatterjee2,chatterjee3,brown2,brown}. In case of one such Ni-Mn-Sn alloys (nominal composition, Ni$_2$Mn$_{1.4}$Sn$_{0.6}$), the rise in $\rho$ around the MPT is about 36\%, which is comparable to 43\% rise in the presently studied GdPd$_2$Bi.   
 
\par
The analysis of our $T$ dependent PXRD data indicates that the sample undergoes structural transition from cubic ($Fm\bar{3}m$) to orthorhombic ($Pmma$) structure [see fig~\ref{xrd}]. For GdPd$_2$Bi, we observe robust Bain distortion ~\cite{sanjaysingh,brown} between cubic and orthorhombic cell parameters: $a_{orth} = \sqrt{2}a_{cub}$, $b_{orth} = a_{cub}$ and $c_{orth} = a_{cub}/\sqrt{2}$. Similar lattice transformation is also observed in case of Co$_2$NbSn, Ni$_2$Mn$_{1.44}$Sn$_{0.56}$ and Ni$_2$Mn$_{1.48}$Sb$_{0.52}$  samples across the MPT~\cite{co2nbsn,nimnsn}. For GdPd$_2$Bi, we observe a 1.5\% change in lattice volume across the transition, although the lattice distortion is significant. MPT is a shear-dominated non-diffusive solid to solid phase transition, where the volume change is small~\cite{chernenko}. Therefore, the observed transition in GdPd$_2$Bi is likely to be martensitic-type. The MPT in GdPd$_2$Bi is clearly first order in nature, which is supported by the phenomenon of phase coexistence obtained from our PXRD data. The further observe thermal hysteresis in our $C_P$ data, which strengthens the first order nature of the MPT.     

\par
Another important feature is the occurrence of large negative MR in the title compound. The MR follows an $H^2$ dependence up to 150 kOe of field and it remains unsaturated. The negative MR is prevalent in the PM phase well above the AFM transition temperature. There are reports of negative MR in several other intermetallic AFM compounds in the PM phase, such as Gd$_2$PdSi$_3$, GdSi, GdNi$_2$Si$_2$, FeSe$_2$~\cite{gd2pdsi3,gdsi,gdni2si2,fese2} and so on. For GdNi$_2$Si$_2$ negative $MR$ is found both in the PM ($T\gtrsim T_N$) state and also below $T_N$ varying almost quadratically with $H$. Such behavior, is not attributed to the  Kondo effect due to the well-localized character of Gd $4f$ electrons, but rather depends on the spin fluctuations in the  Ni $3d$ band~\cite{gdni2si2}. The negative MR close to $T_N$ arises due to the suppression of spin fluctuations by the magnetic field. However, for some  compounds (e.g., Gd$_2$PdSi$_3$, GdSi), considerable MR exists well above $T_N$, and it is attributed to the formation of magnetic-polarons. The applied field aligns these magnetic polaron (local FM clusters) leading to the reduction of spin-dependent scattering~\cite{gdsi}. 
\par
From a theoretical point of view, Usami {\it et al.}~\cite{usami} studied the MR response of AFM metal in the light of  $s-d$ model of electrons. In presence of fluctuating moments from $d$ electrons, the calculation shows a quadratic $H$ dependence of negative MR at low field even in the PM, which eventually saturates at higher fields. 

\par  
In case of GdPd$_2$Bi, finite negative MR is observed for temperature as high as 150 K [see fig.~\ref{res}(d)] and it is quite unlikely to observe antiferromagnetic fluctuations at temperatures much higher than the $T_N$. A polaronic picture can be mooted for GdPd$_2$Bi similar to Gd$_2$PdSi$_3$ or GdSi. However, the magnetic susceptibility of GdPd$_2$Bi shows a perfect Curie-Weiss behavior down to about 15 K, and the data recorded at $H$ = 100 Oe and 1 kOe overlap with each other (see lower inset of fig~\ref{mag}). This does not support the picture of spin fluctuations or  magnetic polarons, as they would lead to the deviation from the Curie-Weiss law. Additionally, had the MR been related to the spin-fluctuation or magnetic polarons, one would expect a saturation or tendency towards saturation at higher fields. However, we observe a robust $H^2$ behaviour even at a field as high as 150 kOe. At such a high field, the spin-fluctuation should get completely suppressed or the polarons would have melted. Therefore, a simple spin-fluctuation or polaronic model may not be appropriate here.

\par
GdPd$_2$Bi shows martensitic transition, and it may contain martensitic variants at low temperature~\cite{variant}. It has been found (specially in case of  ferromagnetic shape memory alloys) that a magnetic field can induce a reverse transition from martensite to austenite leading to negative MR. However, for GdPd$_2$Bi, we found that the martensitic transition temperature is completely insensitive of $H$, and therefore, such phenomenon can be ruled out. 
\par  
A non-saturating positive MR varying quadratically with $H$ is not uncommon. There are several systems having non-trivial topology in their electronic structure, which show nonsaturarting positive MR~\cite{ali,nbp,ru2sn3,ptse2,tasb2}. However, the presently studied compound does not belong to that category.  Saini {\it et al.} recently reported quadratically varying positive MR in the single crystal sample of PtAl~\cite{ptal}. The Fermi surface contains both electron and hole pockets similar to the presently studied compound. It appears that the negative MR in GdPd$_2$Bi does not have a magnetic or structural  origin, and it is possibly connected to the complex nature of the Fermi surface. 
\par
The most fascinating feature here is the presence of distinct thermal hysteresis in the $\rho$ vs $T$ data both in zero field as well as $H$ = 100 kOe.  Thermal hysteresis in $\rho$ vs $T$ data is a common feature in the case of Ferromagnetic shape memory alloys (FSMAs) (such as Co–Ni–Al, Ni-Mn-Sn) that undergo a first order phase transition from austenite (cubic) to martensitic (orthorhombic/hexagonal/tetragonal/monoclinic) phase. Interestingly, similar scenario of martensitic phase transformation occurs for GdPd$_2$Bi.  

\par
Further evidence in support of occurrence of first order transition is seen  from the $C_P$ and $\alpha_S$ versus $T$ data. A broad anomaly is observed in $C_P$ around 160 K. First order martensitic phase transitions often give rise to an anomaly in $C_P$, which is of finite width as a result of the system not being in thermal equilibrium~\cite{co2nbsn}.

\par
The $\alpha_S$ value for GdPd$_2$Bi is negative for $T \geq$ 25 K and positive below it. Interestingly, the sign of $R_H$ is negative indicating electrons to be the dominant charge carrier in GdPd$_2$Bi, which is in line with $\alpha_S$. However, below about 30 K, the $\rho_{xy}$ turns non-linear indicating the significant contribution of holes in addition to electrons in the system. It might be possible that the contribution from hole carriers becomes significant for $T <$ 30 K, which in turn leads to small positive value of $\alpha_S$. The Fermi surface plot also shows that both HT and LT, two types of carriers i.e. electron and hole exist. However, in the LT phase, the hole contribution as a carrier is more prominent than the HT phase.

\section{Summary and Conclusions}
An in-depth study of structure, magnetic and electronic ground state of full Heusler GdPd$_2$Bi compound is presented via x-ray diffraction, magnetization, magnetotransport, thermal transport and heat capacity measurements with further support from DFT calculations. Magnetic study reveal that the compound orders antiferromagnetically at around 7 K. The total energy calculations further confirm that the antiferromagnetic correlation is dominant over the full polarized spin configuration for GdPd$_2$Bi. Magnetotransport study reveals uncharacteristic thermal hysteresis in 100-200 K range, while no corresponding singularities are observed in the magnetic data. In addition, a fairly high quadratic and nonsaturating magnetoresistance has been observed in this compound. We propose that the negative MR is not directly related to the magnetic properties of the material, rather it arises from the complex nature of the Fermi surface with nesting and electron/hole pockets~\cite{puddles}. Detailed structural analysis reveal a martensitic type structural phase transition is likely to be associated with the thermal hysteresis and no gap opening occurs at $E_F$ for both HT and LT phases. The Hall measurement and electronic structure calculations in presence of SOC corroborate complex character of the electronic structure near Fermi surface due to the presence of both electron and hole bands in poor metallic GdPd$_2$Bi. SOC is an important parameter in rare-earth based systems which significantly influences physical properties in a system. Tuning the electronic structure can be possible by altering the rare-earth element in REPd$_2$Bi series. Therefore, our work not only present a thorough study on GdPd$_2$Bi, but also would boost interest for further investigation on the other members of the REPd$_2$Bi family.

\section*{Acknowledgment}
Snehashish Chatterjee, Prabir Dutta and Surasree Sadhukhan would like to acknowledge Univerisity Grants Commission (India), Science and Engineering Research Board (Grant No. PDF/2017/001061), and IIT Goa respectively for their research fellowship. Sudipta Kanungo thanks Department of Science and Technology, Govt. of India for providing an INSPIRE faculty research grant [Grant No. DST/INSPIRE/04/2016/000431; IFA16-MS91]. UGC DAE-CSR, Indore, and Kolkata Centers are duly acknowledged for low-temperature transport and heat-capacity measurements. The authors would like to sincerely acknowledge the Indian beamline facility at Photon Factory, KEK, Tsukuba, Japan for low temperature synchrotron x-ray diffraction measurement.


\end{document}